\magnification=1170 \vsize=8.9truein \hsize=6.5truein \baselineskip=0.6truecm
\parindent=0truecm \parskip=0.2cm \nopagenumbers \font\scap=cmcsc10
\hfuzz=0.8truecm
\font\tenmsb=msbm10
\font\sevenmsb=msbm7
\font\fivemsb=msbm5
\newfam\msbfam
\textfont\msbfam=\tenmsb
\scriptfont\msbfam=\sevenmsb
\scriptscriptfont\msbfam=\fivemsb

\def\hfl#1#2{\smash{\mathop{\hbox to 7mm{\rightarrowfill}}
\limits^{\scriptstyle#1}_{\scriptstyle#2}}}

\def\vfl#1#2{\llap{$\scriptstyle #1$}\left\uparrow
\vbox to 4mm{}\right.\rlap{$\scriptstyle #2$}}

\def\adots{\mathinner{\mkern2mu\raise1pt\hbox{.}
\mkern3mu\raise4pt\hbox{.}\mkern1mu\raise7pt\hbox{.}}}

\def\prof{\vrule depth 3.7pt width 0pt}
\def\sti#1#2#3{\prof{\smash{\hbox{\vtop{\hbox{$#1$}%
\hbox{\raise#2pt\hbox{$\mkern#3mu\tilde{}$}}}}}}}

\null\bigskip\centerline{\bf Integrable lattice equations and their growth
properties}
\vskip 2truecm
\centerline{\scap S. Tremblay}
\centerline{\sl Centre de Recherches Math\'ematiques}
\centerline{\sl et D\'epartement de Physique}
\centerline{\sl Universit\'e de Montr\'eal}
\centerline{\sl C.P. 6128, Succ.~Centre-ville}
\centerline{\sl Montr\'eal, H3C 3J7, Canada}
\bigskip
\centerline{\scap B. Grammaticos}
\centerline{\sl GMPIB, Universit\'e Paris VII}
\centerline{\sl Tour 24-14, 5$^e$ \'etage, case 7021}
\centerline{\sl 75251 Paris, France}
\bigskip
\centerline{\scap A. Ramani}
\centerline{\sl CPT, Ecole Polytechnique}
\centerline{\sl CNRS, UMR 7644}
\centerline{\sl 91128 Palaiseau, France}
\bigskip
\bigskip\bigskip
Abstract
\medskip
In this paper we investigate the integrability of two-dimensional partial
difference equations using
the newly developed techniques of study of the degree of the iterates. We
show that while for generic,
nonintegrable equations, the degree grows exponentially fast, for
integrable lattice equations
the degree growth is polynomial. The growth criterion is used in order to
obtain the integrable
deautonomisations of the equations examined. In the case of linearisable
lattice equations we
show that the degree growth is slower than in the case of equations
integrable through Inverse
Scattering Transform techniques.


\vfill\eject
\footline={\hfill\folio} \pageno=2
The study of integrability of nonlinear evolution equations  has spurred
the development of efficient tools for
 its detection. The ARS [1] conjecture was formulated originally
for partial differential equations and
related integrability to the Painlev\'e property. In the discrete domain
the singularity confinement [2] property was
discovered while studying the lattice KdV equation and the singularities
that can appear spontaneously during the
evolution. The singularity confinement has been a most useful  discrete
integrability criterion in the sense that it is a necessary condition for
lattice equations to be integrable by Inverse Scattering Transform (IST)
methods. However, it
has turned out [3] that
singularity confinement is not sufficient for integrability and thus its
use as an
integrability detector must be subject to particular caution.

Another property of integrable discrete systems, namely the growth of the
degree of the iterates [4],
has, in the long run, proven to be a reliable integrability detector. The
main idea goes back to Arnold [5]  and Veselov [6].
As Veselov summarized it: ``integrability has an essential correlation with
the weak growth
of certain characteristics''.
The characteristic quantity which can be easily obtained and computed for a
rational mapping is the degree of the numerators or denominators of (the
irreducible forms of)
the iterates of some initial condition. (In order to obtain the degree one
must introduce
homogeneous coordinates and compute the homogeneity degree).
Those ideas were refined by Viallet and collaborators [7,8], leading to the
introduction of the notion of algebraic entropy. The latter is defined as
$E=\lim_{n\to \infty} \log(d_n)/n$ where $d_n$ is the degree of the $n$-th
iterate. A generic, nonintegrable, mapping leads to exponential growth
of the degrees of the iterates and thus has a nonzero algebraic entropy,
while an
integrable mapping has zero  algebraic entropy. As we have shown in a previous
work [9],
this is too crude an estimate. The degree growth contains information that
can be
an indication as to the precise integration method to be used  and thus
should be
studied in detail. (At this point, we must stress that, as was already
pointed out
in [8], the degrees of the iterates are {\sl not} invariant under
transformation of
the variables. However the degree {\sl growth} is invariant and
characterises the system at hand).

In  previous  works of ours we have applied the techniques of degree growth
to the
study of one-dimensional mappings [9,10]. A first important conclusion of these
studies was
the confirmation of the singularity confinement  results [11] on the
derivation of
discrete Painlev\'e equations. We have shown that, when singularity
confinement is
used for  the deautonomisation of an integrable autonomous mapping, the
condition
obtained is identical to the one found by requiring nonexponential growth
of the degrees of the iterates. (The terms ``degrees of the iterates'' in
the above sentence and in
the rest of the paper must be understood as the common homogeneity degree
of the numerators and
denominators of their irreducible forms, obtained through the introduction
of the homogeneous
coordinates). This not only confirms the results previously obtained for
discrete Painlev\'e equations,
but also suggests a dual strategy for the study of discrete integrability
based on the combined use of
singularity confinement and study of degree growth. The second result [9]
was that mappings which are
linearisable
are associated to a degree growth slower than the ones integrable through
IST techniques. Thus, the detailed study of the
degree is not
only an indication of integrability but also of the integration method.

In this paper, we apply the techniques of degree growth to two-dimensional
 partial difference equations. We shall show that the main conclusions from
the study of one-dimensional mappings carry over to the two-dimensional
case in a rather straightforward way.

Let us start with the examination of the equation that serves as a
paradigm in all
integrability studies, namely KdV, the discrete form of which is [12,13]:
$$X^{m+1}_{n+1}=X^m_n+{1\over X^{m}_{n+1}} -{1\over X^{m+1}_{n}}.\eqno(1)$$
(Incidentally, this is precisely the equation we have studied in [2], while
investigating the singularity confinement property.) The study of the
degree growth
of the iterates in the case of a 2-dimensional lattice is substantially more
difficult than that of the 1-dimensional case.  It is thus very important
to make the right choices from the outset. Here are the initial conditions we
choose: on the line $m=0$ we take $X^0_n$ of the form $X^0_n=p_n/q$ while on
the line
$n=0$ we choose $X^m_0=r_m/q$ (with $r_0=p_0$). We assign to $q$ and
the $p$'s, $r$'s  the same degree of homogeneity. Then we
compute the iterates of  $X$ using  (1) and calculate the
degree of homogeneity in $p$, $q$, $r$ at the various points of the
lattice. Here is what we find:
$$\matrix{
& \!\!\!\!\!\!\vdots & \!\!\!\!\!\!\vdots & \!\!\!\!\!\!\vdots &
\!\!\!\!\!\!\vdots &
\!\!\!\!\!\!\vdots & \!\!\!\!\!\!\vdots
&\adots \cr
\cr
&          1\quad & 7\quad &  19\quad &  31\quad &  41\quad & 51\quad &
\cdots \cr
\cr
&          1\quad & 5\quad &  13\quad &  19\quad &  25\quad & 31\quad &
\cdots \cr
\cr
&          1\quad & 3\quad &  5\quad &  7\quad &  9\quad & 11\quad & \cdots \cr
\cr
\vfl{m}{}& 1\quad & 1\quad &  1\quad &  1\quad &  1\quad & 1\quad &\cdots \cr
&\hfl{}{n}
}$$

At this point  we must indicate how the analytical expression for the
degree can be
obtained. First we compute several points on the lattice which allow us to
have a
good guess at how the degree behaves.  In the particular case of a
2-dimensional discrete equation relating four points on  an elementary
square like
(1), and with the present choice of initial conditions (and given our
experience on
1-dimensional mappings) we can reasonably surmise that the dominant
behaviour of the
degree will be of the form $d^m_{n}\propto  mn$. Moreover the subdominant
terms
must be symmetric in $m$, $n$ and at most linear. With those indications it is
possible to ``guess'' the expression $d^m_{n}=4mn-2\max(m,n)+1$ (for
$mn\neq 0$)
and subsequently calculate some more points in order to check its validity.
This procedure will be used throughout this paper.

So the lattice KdV equation leads, quite expectedly, to a polynomial growth
in the degrees of the iterates. Let us now turn to the more interesting
question of
deautonomisation. The form (1) of KdV is not very convenient and thus we shall
study its potential form [14]:
$$x^{m+1}_{n+1}=x^m_n+{z_n^m\over x^{m+1}_{n}-x^{m}_{n+1}}.\eqno(2)$$
(The name `potential' is given here in analogy to the continuous case:
the dependent variable $x$ of equation (2) is related to the dependent
variable $X$ of
equation (1) through $x^{m+1}_{n}-x^{m}_{n+1}=X_n^m$ and (1) is recovered
exactly if
$z_n^m$=1). The  deautonomisation we are referring to consists in finding
an explicit
$m,n$ dependence of $z^m_n$ which is compatible with integrability. Let us
first
compute the degrees of the iterates for constant $z$:
$$\matrix{
& \!\!\!\!\!\!\vdots & \!\!\!\!\!\!\vdots & \!\!\!\!\!\!\vdots &
\!\!\!\!\!\!\vdots &
\!\!\!\!\!\!\vdots & \!\!\!\!\!\!\vdots
&\adots \cr
\cr
&          1\quad & 4\quad &  7\quad &  10\quad &  13\quad & 16\quad &
\cdots \cr
\cr
&          1\quad & 3\quad &  5\quad &  7\quad &  9\quad & 11\quad & \cdots \cr
\cr
&          1\quad & 2\quad &  3\quad &  4\quad &  5\quad & 6\quad & \cdots \cr
\cr
\vfl{m}{}& 1\quad & 1\quad &  1\quad &  1\quad &  1\quad & 1\quad &\cdots \cr
&\hfl{}{n}
}$$
The degree  $d^m_{n}$ is given simply by $d^m_{n}=mn+1$. Assuming a generic
$(m,n)$
dependence for $z$ we obtain the following successive degrees:

$$\matrix{
& \!\!\!\!\!\!\vdots & \!\!\!\!\!\!\vdots & \!\!\!\!\!\!\vdots &
\!\!\!\!\!\!\vdots &
\!\!\!\!\!\!\vdots & \!\!\!\!\!\!\vdots
&\adots \cr
\cr
&          1\quad & 4\quad &  10\quad &  20\quad &  35\quad & 56\quad &
\cdots \cr
\cr
&          1\quad & 3\quad &  6\quad &  10\quad &  15\quad & 21\quad &
\cdots \cr
\cr
&          1\quad & 2\quad &  3\quad &  4\quad &  5\quad & 6\quad & \cdots \cr
\cr
\vfl{m}{}& 1\quad & 1\quad &  1\quad &  1\quad &  1\quad & 1\quad &\cdots \cr
&\hfl{}{n}
}$$
We remark readily that the degrees form a Pascal triangle i.e. they are
identical
to the binomial coefficients, leading to an exponential growth at
least on a strip along the diagonal.
The way to obtain an integrable
deautonomisation is to require that the degrees obtained in the autonomous and
nonautonomous cases be identical. The first constraint can be obtained by
reducing the
degree of $x^2_2$ from 6 to 5. As a matter of fact, starting from the
initial conditions
$x^0_n=p_n/q$ , $x^m_0=r_m/q$ (with $r_0=p_0$) we obtain
$x^1_1=(p_1p_0-p_0r_1-z_0^0q^2)/(q(p_1-r_1))$, $x_1^2=Q_3/(qQ_2)$ where
$Q_k$ is a
polynomial of degree $k$, and a similar expression for $x_2^1$. Computing
$x_2^2$ we find
$x_2^2=Q_6/(q(p_1-r_1)Q_4)$. It is impossible for $q$ to divide $Q_6$ for
generic initial
conditions. However, requiring ($p_1-r_1$) to be a factor of $Q_6$ we find
the constraint
$z^{1}_{1}-z^{1}_{0}-z^{0}_{1}+z^0_0=0$.
The relation of this result to singularity confinement is quite easy to
perceive. The
singularity corresponding to $q=0$ is indeed a fixed singularity: it exists
for {\sl all}
$(n,m)$'s where either $n$ or $m$ are equal to zero. On the other hand the
singularity related
to $p_1-r_1=0$ appears only at a certain iteration and is thus movable. The
fact that with the
proper choice of $z_n^m$ the denominator factors out, is precisely what one
expects for
the singularity to be confined.

Requiring that $z$ satisfy
$$z^{m+1}_{n+1}-z^{m+1}_{n}-z^{m}_{n+1}+z^m_n=0\eqno(3)$$
suffices to reduce the degrees of all higher $x$'s to those of
the autonomous case. The solution of (3) is $z^m_n=f(n)+g(m)$ where $f$,
$g$ are two arbitrary  functions. This form of $z^m_n$ is precisely the one
obtained in the analysis of convergence acceleration algorithms [15] using
singularity confinement. The integrability of the nonautonomous form of (2)
(and its
relation to cylindrical KdV) has been discussed by Nagai and Satsuma [16]
in the framework of
the bilinear formalism.

We must point out here that the kind of initial conditions we choose, while
influencing the
specific degrees obtained, do not modify the conclusions on the type of
growth. Let us
illustrate this by choosing for (2) a staircase type of initial conditions
where
$x_n^{-n}=p_n/q$, $x_n^{1-n}=r_n/q$ with the same convention as to the
degrees of $q$ and
the $p$'s, $r$'s (but without the now unnecessary constraint $p_0=r_0$). We
find the
degrees:
$$\matrix{
& \!\!\!\!\!\!\vdots & \!\!\!\!\!\!\vdots & \!\!\!\!\!\!\vdots &
\!\!\!\!\!\!\vdots &
\!\!\!\!\!\!\vdots & \!\!\!\!\!\!\vdots
&\adots \cr
\cr
&          1\quad & 2\quad &  4\quad &  7\quad &  11\quad & 16\quad &
\cdots \cr
\cr
&          1\quad & 1\quad &  2\quad &  4\quad &  7\quad & 11\quad & \cdots \cr
\cr
&           \quad & 1\quad &  1\quad &  2\quad &  4\quad & 7\quad & \cdots \cr
\cr
&           \quad &  \quad &  \underline 1\quad &  1\quad &  2\quad &
4\quad & \cdots \cr
\cr
\vfl{m}{}&  \quad &  \quad &   \quad &  1\quad &  1\quad & 2\quad &\cdots \cr
&\hfl{}{n}
}$$
where the underlined 1 corresponds to the origin. The growth is again
quadratic and
depends only on the sum $N=n+m$ of the coordinates: $d_n^m=1+N(N-1)/2$.

Two more well-known discrete equations can be treated along the same lines.
In the case of the lattice mKdV [14]:
$$x^{m+1}_{n+1}=x^m_n\ {x^{m+1}_{n}-z^m_nx^{m}_{n+1}\over
z^m_nx^{m+1}_{n}-x^{m}_{n+1}}\eqno(4)$$
we obtain for constant $z$ the same degree growth, $d^m_{n}=mn+1$, as for
the potential
lattice KdV.
If we assume now a generic $z$ we find the degrees:
$$\matrix{
& \!\!\!\!\!\!\vdots & \!\!\!\!\!\!\vdots & \!\!\!\!\!\!\vdots &
\!\!\!\!\!\!\vdots &
\!\!\!\!\!\!\vdots & \!\!\!\!\!\!\vdots
&\adots \cr
\cr
&          1\quad & 4\quad &  13\quad &  32\quad &  65\quad &   \quad &
\cdots \cr
\cr
&          1\quad & 3\quad &  7\quad &  13\quad &  21\quad & 31\quad &
\cdots \cr
\cr
&          1\quad & 2\quad &  3\quad &  4\quad &  5\quad & 6\quad & \cdots \cr
\cr
\vfl{m}{}& 1\quad & 1\quad &  1\quad &  1\quad &  1\quad & 1\quad &\cdots \cr
&\hfl{}{n}
}$$
The degrees obey the recursion  $d^{m+1}_{n+1}=d^{m+1}_{n}+d^{m}_{n+1}+d^m_n-1$
leading to an exponential growth with asymptotic ration (1+$\sqrt 2$).
Requiring the
degree of $x^2_2$ to be 5 instead of 7 we find the condition
$$z^{m+1}_{n+1}z^m_n-z^{m+1}_{n}z^{m}_{n+1}=0\eqno(5)$$
with solution $z^m_n=f(n)g(m)$. This condition is sufficient for the
degrees of the
nonautonomous case to coincide with those of the autonomous one. It is also
precisely the
one obtained in [15] using the singularity confinement condition. We
believe that the
Nagai-Satsuma approach [16] for the construction of double Casorati
determinant solutions can
be extended to the case of the nonautonomous lattice modified-KdV.

The discrete sine-Gordon equation [17,18]:
$$x^{m+1}_{n+1}x^m_n={1+z^m_nx^{m+1}_{n}x^{m}_{n+1}\over
x^{m+1}_{n}x^{m}_{n+1}+z^m_n}\eqno(6)$$
in the autonomous case where $z$ is a constant leads to the degree pattern:
$$\matrix{
& \!\!\!\!\!\!\vdots & \!\!\!\!\!\!\vdots & \!\!\!\!\!\!\vdots &
\!\!\!\!\!\!\vdots &
\!\!\!\!\!\!\vdots & \!\!\!\!\!\!\vdots
&\adots \cr
\cr
&          1\quad & 5\quad &  9\quad &  13\quad &  16\quad & 19\quad &
\cdots \cr
\cr
&          1\quad & 4\quad &  7\quad &  9\quad &  11\quad & 13\quad &
\cdots \cr
\cr
&          1\quad & 3\quad &  4\quad &  5\quad & 6\quad & 7\quad & \cdots \cr
\cr
\vfl{m}{}& 1\quad & 1\quad &  1\quad &  1\quad &  1\quad & 1\quad &\cdots \cr
&\hfl{}{n}
}$$
It can be represented by $d^{m}_{n}=mn+\min(m,n)+1$.
In the nonautonomous case of generic $z$ we obtain the sequence of
degrees:
$$\matrix{
& \!\!\!\!\!\!\vdots & \!\!\!\!\!\!\vdots & \!\!\!\!\!\!\vdots &
\!\!\!\!\!\!\vdots &
\!\!\!\!\!\!\vdots & \!\!\!\!\!\!\vdots
&\adots \cr
\cr
&          1\quad & 5\quad &  19\quad &  49\quad & 96 \quad &   \quad &
\cdots \cr
\cr
&          1\quad & 4\quad &  11\quad &  19\quad &  29\quad & 41\quad &
\cdots \cr
\cr
&          1\quad & 3\quad &  4\quad &  5\quad & 6\quad & 7\quad & \cdots \cr
\cr
\vfl{m}{}& 1\quad & 1\quad &  1\quad &  1\quad &  1\quad & 1\quad &\cdots \cr
&\hfl{}{n} }$$
obeying the relation
$d^{m+1}_{n+1}=d^{m+1}_{n}+d^{m}_{n+1}+d^m_n-(1-\delta^m_n)$ leading
again to exponential growth. The condition for a growth identical to that
of the autonomous case is the same as (5). Thus equation (6) introduces a
nonautonomous extension of the
lattice sine-Gordon equation. (We intend to return to a study of its
properties in some future work).
We must point out that in the continuous limit, this nonautonomous form
goes over to
$w_{x,t}=f(x)g(t)\sin w$. This explicit $x$ and $t$ dependence can be absorbed
through a redefinition of the independent variables leading to the standard,
autonomous, sine-Gordon, but no such gauge exists in the discrete case.

We now turn to two discrete equations which are particular in the sense
that they are not
integrable through IST techniques but rather  through direct linearisation.
The first
is the discrete Liouville equation [19]:
$$x^{m+1}_{n+1}x^m_n=x^{m+1}_{n}x^{m}_{n+1}+z^m_n.\eqno(7)$$
If we assume that $z$ is a constant we obtain the following degree pattern:
$$\matrix{
& \!\!\!\!\!\!\vdots & \!\!\!\!\!\!\vdots & \!\!\!\!\!\!\vdots &
\!\!\!\!\!\!\vdots &
\!\!\!\!\!\!\vdots & \!\!\!\!\!\!\vdots
&\adots \cr
\cr
&          1\quad & 4\quad &   5\quad & 6\quad & 7\quad & 8\quad & \cdots \cr
\cr
&          1\quad & 3\quad &  4\quad &  5\quad & 6\quad & 7\quad & \cdots \cr
\cr
&          1\quad & 2\quad &  3\quad &  4\quad &  5\quad & 6\quad & \cdots
\cr
\cr
\vfl{m}{}& 1\quad & 1\quad &  1\quad &  1\quad &  1\quad & 1\quad &\cdots \cr
&\hfl{}{n} }$$
By inspection we find $d^{m}_{n}=m+n$. This result is not at all
astonishing. As we have
shown in [9], the degree growth of linearisable mappings is slower than that
of the IST
integrable ones. The same feature appears again here. The deautonomisation
of (7) can
proceed along the same lines as previously. For generic $z_n^m$, the
degrees are
organised in a Pascal triangle and thus the growth is exponential. The
condition for the
growth to be identical to that
of the autonomous case is again (5) and thus $z^m_n=f(n)g(m)$. However,
this nonautonomous extension is trivial: it can be obsorbed through a
simple gauge
transformation. Indeed,
putting $x=\phi X$ where $\phi=\alpha(n)\beta(m)$ with
$f(n)=\alpha(n)\alpha(n+1)$,
$g(m)=\beta(m)\beta(m+1)$ we can reduce equation (7) to one where $z\equiv1$.

Finally, we analyse the discrete Burgers equation [19]:
$$x^{m+1}_{n}=x^m_n{1+z^m_n x^{m}_{n+1}\over 1+z^m_n x^{m}_{n}}.\eqno(8)$$
When $z$ is a constant we find $d^m_n=m+1$. (Notice that contrary to all
the previous
examples, in the case of Burgers equation $m$ and $n$ do not play the same
role and
thus a $d^m_n$ that is not symmetric in $m$, $n$ is not surprising).  For a
generic
$z^m_n$, we find $d^m_n=2^m$, a manifestly exponential growth. The
condition for the
degree to grow like $m+1$ is just
$$z^{m}_{n+1}-z^m_n=0\eqno(9)$$
i.e. $z^{m}_{n}=g(m)$. This leads to a nonautonomous extension of the
lattice Burgers equation.
Moreover this extension cannot be removed by a gauge. On the other hand,
this extension is perfectly
compatible with linearisability. Indeed, putting
$x^{m}_{n}=X^{m}_{n+1}/X^{m}_{n}$ we can reduce it
to the linear equation:
$$X^{m+1}_{n}=f(m)(X^m_n+g(m) X^{m}_{n+1})\eqno(10)$$
where $f$ is arbitrary and can be taken equal to unity. This nonautonomous
extension is
just a special case of the more general discrete Burgers:
$$x^{m+1}_{n}=x^m_n{\alpha^m_n+\beta^m_n x^{m}_{n+1}\over 1+\gamma^m_n
x^{m}_{n}}\eqno(11)$$
which can be linearised through $x^m_n=\phi^m_nX^{m}_{n+1}/X^m_n$
to $X^{m+1}_{n}=\psi^m_n(X^m_n+\gamma^m_n\phi^m_nX^{m}_{n+1}) $
provided $\beta^m_n=\alpha^m_n\gamma^{m}_{n+1}$ and $\alpha$, $\phi$ and
$\psi$ are related through
$\alpha^m_n\psi^m_n\phi^m_n=\psi^{m}_{n+1}\phi^{m+1}_{n}$.
We must point out here that the continuous Burgers equation also does
possess  a
nonautonomous extension. It is straightforward to show that if $\phi$ is a
solution
of the equation
$\phi_t=\phi^2\phi_{xx}$ then the nonautonomous Burgers $u_t=\phi^2
u_{xx}+2\phi uu_{x}$ can be
linearised to
$v_t=\phi^2v_{xx}$ through the Cole-Hopf transformation $u=\phi v_x/v$.

In this paper, we have applied the method of the slow degree growth to the
study of the integrability
of partial difference equations. Our study has focused on well-known
integrable lattice
equations for which we have tried to provide nonautonomous forms. We have
shown that using
degree-growth methods it is possible to obtain integrable nonautonomous
forms for most of the
equations studied, and confirmed results previously obtained through the
singularity
confinement  method. In the case of linearisable lattice equations, our
results are the
logical generalisation of the ones obtained for 1-dimensional mappings: the
linearisable
mappings have a degree growth that is slower than the one of the
IST-integrable discrete
equations. Our estimate of the degree growth was based on the direct
computation of the degree
for successive iterations and obtaining a fit of some analytical expression
confirmed by
subsequent iterations. It would be interesting, of course, to provide a
rigorous proof of the
degree growth following, for instance, the methods of [20]. However, this
has not yet been
carried through even for one-dimensional, nonautonomous mappings that are
integrable through
spectral methods. On the other hand, the proof of the degree growth for the
cases where the
equations are linearisable looks more tractable and we intend to address
this question for both
the one-and two-dimensional cases in some future work.

The fact that we were able, through the adequate choice of initial data, to
perform these
calculations without being overwhelmed by their size is an indication of
the usefulness of our
approach. The study of degree growth, perhaps coupled with singularity
confinement in the dual
strategy we sketched in [10], can be a precious tool for the detection of
integrability of
multidimensional discrete systems. The interest of this method is not only
that it can be used
as a detector of new integrable lattice systems but also that it can
furnish an indication as
to the precise method of their integration.
\bigskip
{\scap Acknowledgments}
\smallskip
S. Tremblay acknowledges a scholarship from Centre de Cooperation
Internationale Franco-Qu\'ebecois.
\bigskip
{\scap References}
\medskip
\item{[1]} M.J. Ablowitz, A. Ramani and H. Segur, Lett. Nuov. Cim. 23
(1978) 333.
\item{[2]} B. Grammaticos, A. Ramani and V. Papageorgiou, Phys. Rev. Lett.
67 (1991) 1825.
\item{[3]} J. Hietarinta and C. Viallet, Phys. Rev. Lett. 81, (1998) 325.
\item{[4]} M.P. Bellon, J.-M. Maillard and C.-M. Viallet, Phys. Rev. Lett.
67 (1991) 1373.
\item{[5]} V.I. Arnold, Bol. Soc. Bras. Mat. 21 (1990) 1.
\item{[6]} A.P. Veselov, Comm. Math. Phys. 145 (1992) 181.
\item{[7]} G. Falqui and C.-M. Viallet, Comm. Math. Phys. 154 (1993) 111.
\item{[8]} M.P. Bellon and C.-M. Viallet, Comm. Math. Phys. 204 (1999), 425.
\item{[9]} A. Ramani, B. Grammaticos, S. Lafortune and Y. Ohta, J. Phys. A
33 (2000) L287.
\item{[10]} Y. Ohta, K.M. Tamizhmani, B. Grammaticos and A. Ramani, Phys.
Lett. A. 262 (1999) 152.
\item{[11]} A. Ramani, B. Grammaticos and J. Hietarinta, Phys. Rev. Lett.
67 (1991) 1829.
\item{[12]} R. Hirota, J. Phys. Soc. Japan 43 (1977) 1424.
\item{[13]} V. Papageorgiou, F.W. Nijhoff and H. Capel, Phys. Lett. A 147
(1990) 106.
\item{[14]} H. Capel, F.W. Nijhoff and V. Papageorgiou, Phys. Lett. A 155
(1991) 337.
\item{[15]} V. Papageorgiou, B. Grammaticos and A. Ramani, Phys. Lett. A
179 (1993) 111.
\item{[15]} A. Nagai and J.Satsuma, Phys. Lett. A 209 (1995) 305.
\item{[17]} R. Hirota, J. Phys. Soc. Japan 43 (1977) 2079.
\item{[18]} A. Bobenko, M. Bordemann, C. Gunn and U. Pinkall, Comm. Math.
Phys. 158 (1993) 127.
\item{[19]} R. Hirota, J. Phys. Soc. Japan 46 (1979) 312.
\item{[20]} M.P. Bellon, Lett. Math. Phys. 50 (1999) 79.
\end